\documentstyle[aps,12pt,cite]{revtex}

\begin{document}
\draft
\newcommand{\dis}{\displaystyle}
\newcommand{\expon}{\rm e}
\newcommand{\beq}{\begin{equation}}
\newcommand{\eeq}{\end{equation}}

\title{ Analytic Expression for Exact Ground State Energy
Based on  an Operator Method for a Class of
Anharmonic Potentials} 
\author{}
\author{}
\author{ L. C. Kwek, Yong Liu, C. H. Oh \footnote{Email Address:
phyohch@nus.edu.sg}
and Xiang-Bin Wang } 
\address{ Physics Department, National
University of Singapore, \\ Kent Ridge, Singapore, 119260}
\maketitle

\bigskip
\vspace{3.5cm}

\begin{abstract}
A general procedure based on shift operators is formulated to deal
with anharmonic potentials.  It is possible to extract the ground
state energy analytically using our method provided certain
consistency relations are satisfied. Analytic expressions for the
exact ground state energy have also been derived specifically for
a large class of the one-dimensional oscillator with cubic-quartic
anharmonic terms. Our analytical results can be used to check the
accuracy of existing numerical methods, for instance the method of
state-dependent diagonalization. Our results also agree with the
asymptotic behavior in the divergent pertubative expansion of
quartic harmonic oscillator.
\end{abstract}

\newpage

\section{Introduction}

The operator method has been widely used as an elegant analytical
tool in quantum mechanics \cite{raab,cooper} and quantum
statistics \cite{yang} for studying exactly solvable models. Since
the publication of a remarkable paper by Delbecq and Quesne
\cite{dq}, the operator method has been extended to many other
possible physical applications and realizations
\cite{fgn,balan,gms,dnnr,dgrs,acin,chen1,chen2,kwek}. Indeed, the
generalization of the notion of ``shift operators" or ``ladder
operators" as a spectrum generating algebra  can be effectively
studied both mathematically and physically in terms of ``nonlinear
algebra" introduced by Delbecq and Quesne\cite{dq}. In fact, Chen
et al \cite{chen2} has recently reproduced the results of Bethe
Ansatz \cite{bethe,ft,kre} for the XXX model through the shift
operator method. A general observation in these works appears to
be a possibility of treating the well-known technique of the
algebraic Bethe Ansatz for quantum many-body problem as a special
case of the shift operator method. Thus, all previous works
evaluated using the algebraic Bethe Ansatz should hopefully be
possible using a nonlinear algebraic method via shift operators. A
further implication of this notion is the emergence of an
underlying Yangian algebra. Furthermore, the factorization method
employed in supersymmetric quantum mechanics (SQM)\cite{cks} can
also be deemed as a special case of the nonlinear Delbecq-Quesne
algebra.

In many realizations of physical models, the operator method can
provide us with an alternative and a clearer picture regarding the
analytical determination of energy eigenstates and eigenvalues.
Specifically, it has been found to be a useful tool in the
solution of bound state problems such as the string postulate in
Bethe Ansatz or the raising and lowering of energy levels in SQM.
As in Bethe Ansatz and SQM, knowledge of a certain reference
state, like the highest weight state in Bethe Ansatz or the ground
state ansatz for the superpartner potential in SQM, in the
operator method\cite{raab} permits the analysis and determination
of the full spectrum of a physical system. Moreover, as mentioned
in ref. \cite{cooper}, the knowledge of the ground state wave
function also determines the potential and hence the Hamiltonian
of a system up to a constant value. However, in principle given
the Hamiltonian of a physical system, surely it should be possible
to determine all the energy levels. In this paper, we show that
provided certain consistency relations are satisfied, the entire
spectrum of a given Hamiltonian can be determined by shift
operator method even if there is no prior knowledge of a reference
state.

To illustrate the power of the shift operator method, we apply the
technique to one-dimensional oscillator with anharmonic
potentials. The study of anharmonic potential \cite{cbw,bend,keun}
has always been an exciting and interesting field due to its broad
applications in quantum field theory \cite{ijk,wu3}, nuclear
models, atomic and molecular physics \cite{czb,cpp,cag}, condensed
matter physics\cite{ssc,spr}, statistical physics and chemical
physics \cite{acs}. Indeed, numerous numerical methods including
renormalized strong coupling expansion, perturbation expansion,
supersymmetric quantum mechanics, WKB, iteration based on the
generalized Bloch equation, state-dependent diagonalization, Hill
determinant method, phase-integral approach, iterative Bogoliubov
transformations, eigenvalue moment method, perturbative-variation,
and algebraic method have been proposed to investigate these
anharmonic potentials
\cite{rncm,lo,idfl,idfk,rjjr,hsrt,scds,ia,sdcb,yzdm,byms,hmos,ejw,hacr,fadb,bisb,datt}.
However, due to its inherent intractability, no analytic solution
for energy spectrum and eigenstates has been obtained so far
except in some special cases. Indeed, for the first time, we have
partially resolved some of these technical issues and obtained
analytic solutions for a wide class of anharmonic potentials under
certain conditions.

It is instructive to see how SQM can be regarded as a special case
of the nonlinear algebraic operator method. Before we proceed
further to show this point, we first recall some relevant
definitions in operator method and nonlinear algebra
\cite{raab,dq,chen1,kwek}. Let $\hat{H}$ be an observable
satisfying the eigenfunction equation $\hat{H} |\psi>=E|\psi>$.
The operators $L^{\pm}$ satisfying \beq \lbrack \hat{H}, L^{\pm }
\rbrack=L^{\pm} f^{\pm}(\hat{H}), \label{eq1} \eeq are called
``shift operators" of $\hat{H}$. In Eq.(\ref{eq1}),
$f^{\pm}(\hat{H})$ are real functions of $\hat{H}$ so that the
values of its action on eigenstates of $H$ are reals. Thus, these
values can be interpreted as energy gaps if $\hat{H}$ is the
Hamiltonian of the system. Nevertheless, it is not necessary for
the mutually adjoint condition $$ (L^+)^\dagger=L^- $$ to hold and
there is generally no constraint on the commutation relation
between $L^+$ and $L^-$, i.e. $\lbrack L^+, L^-\rbrack$. However,
if the mutually adjoint condition is satisfied, we obtained the
``nonlinear algebra" \cite{dq} defined by the relations
\begin{eqnarray} \lbrack \hat{H}, L^+ \rbrack & = & f(\hat{H})
L^+, \nonumber \\
 \lbrack \hat{H}, L^- \rbrack & = &-L^- f(\hat{H}) , \nonumber \\
\lbrack L^-, L^+  \rbrack & = & g(\hat{H}) . \nonumber
\end{eqnarray} A simple rearrangement of the above relations gives
\begin{eqnarray} (\hat{H}-f(\hat{H}))L^+ & = & L^+ \hat{H}, \nonumber \\
 L^-
(\hat{H}-f(\hat{H})) & = & \hat{H}L^-. \label{neweq}\end{eqnarray}
It is interesting to compare Eq.(\ref{neweq}) with the analogous
relations in SQM \cite{cks}
\begin{eqnarray} \hat{H}_1 Q^+ & = &Q^+ \hat{H}_2\nonumber \\ Q^-
\hat{H}_1 & = & \hat{H}_2 Q^-\nonumber \\ Q^- & = & (Q^+)^\dagger
\end{eqnarray} and identify
\begin{eqnarray} Q^+ \sim L^+  \hspace{1cm} & Q^- \sim L^-
\nonumber \\ \hat{H}_1 \sim \hat{H}-f(\hat{H})\hspace{0.5cm} &
\hat{H}_2 \sim \hat{H}.
\end{eqnarray}
Hence, one can always regard SQM as a specific type of nonlinear
algebraic method.

The purpose of this paper is to describe the operator method
\cite{raab,chen1,chen2,kwek} and apply it to study the
one-dimensional oscillator with anharmonic potentials. In section
\ref{sect2}, we show how a shift operator method can be formulated
using a simple instructive example of the harmonic oscillator. The
general procedures to solve anharmonic potentials are presented in
section \ref{sect3}. In section \ref{sect4}, we solve the relevant
equations for the one-dimensional oscillator with cubic-quartic
potential and show that we can get the exact analytical expression
for the ground state energy provided certain consistency relations
are satisfied. We then compare our analytical result with some
numerical results in section \ref{sect5}. Our analytical
expressions can be used to verify the accuracy of existing
numerical results. We conclude with some brief remarks in section
\ref{sect6}.

\section{Applying Operator Method to the Harmonic Oscillator}\label{sect2}
In this section, we apply the operator method to the harmonic
oscillator. The Hamiltonian of the harmonic oscillator is given by
\begin{equation}
{\hat H}=\frac{1}{2}( \hat{p}^2+\hat{x}^2).
\end{equation}
Together with the well-known commutation relation between
$\hat{x}$ and $\hat{p}$, i.e. $[\hat{x}, \hat{p}]=i$ ($\hbar=1$),
we get
\begin{eqnarray}
\label{l1} \lbrack \hat{ H}, \hat{ x} \rbrack & =& -i \hat
{p}\nonumber \\ \lbrack \hat{ H}, \hat {p }\rbrack &= &  i \hat
{x}.
\end{eqnarray}
Using the notation $$ \lbrack \hat H,  (\hat {x}, \hat{p}) \rbrack
: = (  \lbrack \hat{H}, \hat{x} \rbrack, \lbrack \hat{H}, \hat{p}
\rbrack  ), $$ Eq.(\ref{l1}) can also be rewritten as
\begin{eqnarray}
\lbrack \hat H, (\hat {x},  \hat{p})\rbrack & = & (\hat{x},
\hat{p}) \left(
\begin{array}{cc}
0 & i\\ -i & 0
\end{array}
\right), \label{l2}\\ & \equiv &  (\hat{x}, \hat{p})  M
\end{eqnarray}
where, the matrix, $M$, is called the ``coefficient matrix". To
diagonalize the coefficient matrix, we look for a transformation,
$U$, such that $M= U D U^{-1}$ where $D$ is a diagonal matrix. In
general, provided all the entries of $U$ commute with Hamiltonian
$\hat{H}$, the entries of the matrix $U$ are functions of $c$
numbers, the Hamiltonian or any conserved quantity $\hat{I}$ of
the system. Thus, we have

\begin{eqnarray}
\lbrack \hat H, (\hat {x},  \hat{p}) U(c, \hat{H}, \hat{I})
\rbrack  &= & (\hat{x}, \hat{p}) U(c, \hat{H}, \hat{I})U^{-1}(c,
\hat{H}, \hat{I}) \left(
\begin{array}{cc}
0 & i\\ -i & 0
\end{array}
\right)U(c, \hat{H}, \hat{I})\nonumber\\ &=& (\hat{x}, \hat{p})
U(c, \hat{H}, \hat{I}) \left(
\begin{array}{cc}
\lambda_1(c, \hat{H}, \hat{I}) &0\\ 0& \lambda_2(c, \hat{H},
\hat{I})
\end{array}
\right )
\end{eqnarray}
 It is easy to see that $$
U=\frac{1}{\sqrt{2}}\left(
\begin{array}{cc}
1 & 1\\ -i& i
\end{array}
\right), $$ so that Eq.(\ref{l2}) can be rewritten as
\begin{eqnarray}
& & \lbrack \hat H, (\hat {x}, \hat{p})\frac{1}{\sqrt{2}}\left(
\begin{array}{cc}
1 & 1\\ -i& i
\end{array}
\right)  \rbrack \nonumber \\ &  = & (\hat{x}, \hat{p})
\frac{1}{\sqrt{2}}\left(
\begin{array}{cc}
1 & 1\\ -i& i
\end{array}
\right) \left(
\begin{array}{cc}
1 & 0\\ 0 & -1
\end{array}
\right). \label{l3}
\end{eqnarray}
Identifying $$ (\hat{ a}^+, \hat{ a})= (\hat{x}, \hat{p})
\frac{1}{\sqrt{2}}\left(
\begin{array}{cc}
1 & 1\\ -i& i
\end{array}
\right), $$ we obtain $$
\begin{array}{ccc}
\lbrack \hat{ H}, \hat{ a}^+ \rbrack & =& \hat{a}^+ \\ \lbrack
\hat{ H}, \hat {a }\rbrack &= &  -\hat {a}.
\end{array}.
$$ The ``creation" and the ``annihilation" operators are thus
obtained naturally. Following the literature, we can also call
them ``shift" or ``ladder" operators. Although, the application of
the shift operator to harmonic oscillator is very simple, it can
still serve an illustrative example.

For the harmonic oscillator,  a closed algebra is obtained. This
closure in the algebra enables the whole spectrum of the system to
be determined using only one pair of shift operators, in which one
of the operator raises while the other lowers the energy levels.
In general, the existence of a closed algebra is not assured, and
so, we usually have to deal with an unclosed algebra with
infinitely many shift operators. Thus, we have to use infinitely
many shift operators pairs to generate the whole spectrum.
Moreover, corresponding to each energy level, we have a pair of
distinct shift operators in which one raises while the other
lowers the energy level.

It is interesting to note that, the eigenvalues of the coefficient
matrix correspond to the amount of shifted energy associated with
the various shift operators. And, except for some special
Hamiltonians, the coefficient matrix, its eigenvalues and the
transformation matrix, are all functions of the Hamiltonian. In
nonlinear algebraic method, the energy shifts are no longer
uniform and the gap between any two adjacent energy levels is
related to their relative positions in the spectrum. Thus, we find
that shift operator method can provide us with a clearer physical
picture for the mapping of the energy level in a physical system.

Note that we can only extract the energy gaps rather than the
energy levels. Moreover, we have no information regarding to the
energy of the ground state\cite{raab}. However, in some cases, it
is possible to determine the exact energy of the ground state.
This last result constitutes the gist of one of the most important
aspect of our paper.

\section{Operator Method to Anharmonic Oscillator: the General
Procedure}\label{sect3}

In this section, we confine ourselves to the one-dimensional
harmonic oscillator in which the anharmonic potential contains
cubic and quartic terms. Nevertheless, the procedure developed
here is very general and can be applied to other types of
anharmonic potentials. The Hamiltonian is given by
\begin{equation}
\hat{H}= - \frac{d^2}{dx^2}+\alpha \hat{x}^2+\beta \hat{
x}^3+\gamma \hat{x}^4,
\end{equation}
and it is easy to show that

\begin{eqnarray}
\label{l4} \lbrack  \hat{H},  \hat{x}^n  \rbrack & =& - 2 n
\hat{x}^{n-1}\frac{d}{dx}- n (n-1) \hat{x}^{n-2}\\ \lbrack
\hat{H},  \hat{x}^n \frac{d}{dx} \rbrack & =& - n (n-1)
\hat{x}^{n-2} \frac{d}{dx}+2 n \hat{x}^{n-1} \hat{H}\nonumber\\
\label{l5} && -2 \alpha (n+1) \hat{x}^{n+1}-2 \beta
(n+\frac{3}{2}) \hat{x}^{n+2} -2 \gamma (n+2) \hat{x}^{n+3}.
\end{eqnarray}

Rewriting  Eqs.(\ref{l4},\ref{l5}) into matrix form, we have
\begin{eqnarray} \label{l6} \lbrack  \hat{H},  ( \hat{x},
\hat{x}^2, \hat{x}^3, \cdots, \hat{x}^n, \cdots )  \rbrack & =& (
\hat{x}, \hat{x}^2, \hat{x}^3, \cdots, \hat{x}^n, \cdots )
M_1\nonumber\\ &&+ (\frac{d}{dx}, x \frac{d}{dx}, x^2
\frac{d}{dx}, \cdots,  x^n \frac{d}{dx}, \cdots   ) N_1+L_1\\
\label{l7} \lbrack  \hat{H}, (\frac{d}{dx}, x \frac{d}{dx}, x^2
\frac{d}{dx}, \cdots,  x^n \frac{d}{dx}, \cdots   ) \rbrack & =& (
\hat{x}, \hat{x}^2, \hat{x}^3, \cdots, \hat{x}^n, \cdots )
M_2\nonumber\\ &&+ (\frac{d}{dx}, x \frac{d}{dx}, x^2
\frac{d}{dx}, \cdots,  x^n \frac{d}{dx}, \cdots   ) N_2+L_2
\end{eqnarray}
where \beq M_1=\left(
\begin{array}{cccccccc}
0 & 0 & -6 & 0 & 0 &\cdots&\cdots&\cdots\\ 0 & 0 & 0 & -12 & 0
&\cdots &\cdots &\cdots \\ 0 & 0 & 0 & 0 & -20 &\cdots &\cdots
&\cdots \\ \vdots & \vdots & \vdots & \vdots & \vdots & \ddots &
\vdots & \vdots \\ 0 &0 &0 &0 &0 &\cdots &-n(n-1)& \cdots \\
\vdots & \vdots & \vdots & \vdots & \vdots & \vdots & \vdots &
\ddots
\end{array}
\right),\eeq \beq N_1=\left(
\begin{array}{cccccc}
-2 & 0 & 0 & \cdots&\cdots&\cdots\\ 0 & -4 & 0 & \cdots
&\cdots&\cdots  \\ 0 & 0 & -6 & \cdots &\cdots&\cdots  \\ \vdots &
\vdots & \vdots & \ddots & \vdots& \vdots  \\ 0 &0 &0 & \cdots &-2
n & \cdots \\ \vdots & \vdots & \vdots & \vdots & \vdots & \ddots
\end{array}
\right), \eeq \beq L_1=(0, \;\; -2, \;\; 0, \;\; 0, \;\; \cdots,
\;\; 0, \;\; \cdots) \eeq \beq M_2=\left(
\begin{array}{cccccccc}
-2 \alpha & 0 & 4 \hat{H} & 0 & 0 & & \cdots & \cdots\\ -3\beta &
-4 \alpha & 0 & 4 \hat{H} & 0 && \cdots & \cdots \\ -4 \gamma &
-5\beta & -6\alpha & 0 & 8 \hat{H} && \cdots & \cdots \\ 0 &
-6\gamma & -7\beta & -8 \alpha & 0 & \ddots & \cdots & \cdots \\
0& 0& -8 \gamma & -9\beta & -10 \alpha & \ddots & 2n \hat{H} &
\cdots \\ 0& 0& 0& -10 \gamma & -11 \beta & \ddots & 0 & \cdots \\
0& 0& 0& 0& -12 \gamma &  \ddots & -2(n+1)\alpha & \cdots \\ 0  &
0 & 0 & 0 & 0 & \ddots & -2(n+\frac{3}{2})\beta & \cdots \\ 0  & 0
& 0 & 0 & 0 & \ddots & -2(n+2)\gamma & \cdots \\ \vdots & \vdots &
\vdots & \vdots & \vdots &  & \vdots & \vdots
\end{array}
\right),\eeq \beq N_2=\left(
\begin{array}{cccccccc}
0 & 0 &-2 & 0 & 0& &\cdots &\cdots\\ 0 & 0 & 0 & -6 &0 &&\cdots
&\cdots  \\ 0 & 0 & 0 & 0 & -12 &&\cdots &\cdots \\ \vdots &
\vdots & \vdots & \vdots & \vdots& \ddots  &\vdots &\vdots \\ 0 &
0 & 0 & 0 & 0 & \cdots  &-n(n+1) &\cdots \\ \vdots & \vdots &
\vdots & \vdots & \vdots& &\vdots &\vdots
\end{array}
\right), \eeq and \beq L_2=(0, \;\; 2\hat{H}, \;\; 0, \;\; 0, \;\;
\cdots, \;\; 0, \;\; \cdots)\eeq It is interesting to note that
for other types of potentials, all matrices except for $M_2$ are
exactly the same.

From Eqs.(\ref{l6},\ref{l7}), we find \beq
\begin{array}{lll}
\lbrack  \hat{H},  ( \hat{x}, \hat{x}^2, \hat{x}^3, \cdots,
\hat{x}^n, \cdots ) R + (\frac{d}{dx}, x \frac{d}{dx}, x^2
\frac{d}{dx}, \cdots,  x^n \frac{d}{dx}, \cdots   ) S \rbrack  \\
=( \hat{x}, \hat{x}^2, \hat{x}^3, \cdots, \hat{x}^n, \cdots ) (
M_1 R+M_2 S) +(\frac{d}{dx}, x \frac{d}{dx}, x^2 \frac{d}{dx},
\cdots,  x^n \frac{d}{dx}, \cdots   ) (N_1 R+N_2 S) +(L_1 R+L_2 S)
\end{array}.
\eeq Thus, if the following relations exist
\begin{eqnarray}
\label{l10} M_1 R+M_2 S&=&R T \\ \label{l11} N_1 R+N_2 S &=& S T
\end{eqnarray}
then we have
\begin{equation}
\label{l9}
\begin{array}{l}
\lbrack  \hat{H},  ( \hat{x}, \hat{x}^2, \hat{x}^3, \cdots,
\hat{x}^n, \cdots ) R + (\frac{d}{dx}, x \frac{d}{dx}, x^2
\frac{d}{dx}, \cdots,  x^n \frac{d}{dx}, \cdots   ) S \rbrack  \\
=\{ ( \hat{x}, \hat{x}^2, \hat{x}^3, \cdots, \hat{x}^n, \cdots ) R
+ (\frac{d}{dx}, x \frac{d}{dx}, x^2 \frac{d}{dx}, \cdots,  x^n
\frac{d}{dx}, \cdots   ) S \}T +(L_1 R+L_2 S)
\end{array}.
\end{equation}
To diagonalize the coefficient matrix $T$, let us suppose that $U$
is the transformation matrix needed. The eigenvalues and the
corresponding shift operators can therefore be written as
\begin{equation}
\Lambda=U^{-1}TU=\mbox{\rm diag}(\lambda_1, \lambda_2, \lambda_3,
\cdots, \lambda_n, \cdots )
\end{equation}
and
\begin{equation}
\begin{array}{l}
(\hat{A}_1, \hat{A}_2, \hat{A}_3, \cdots, \hat{A}_n, \cdots )\\
=\{ ( \hat{x}, \hat{x}^2, \hat{x}^3, \cdots, \hat{x}^n, \cdots ) R
+ (\frac{d}{dx}, x \frac{d}{dx}, x^2 \frac{d}{dx}, \cdots,  x^n
\frac{d}{dx}, \cdots   ) S \}U +(L_1 R+L_2 S)U \Lambda^{-1}
\end{array}
\end{equation}
respectively. In fact, Eq.(\ref{l9}) can be written into a more
succinct form as \beq \lbrack  \hat{H},  ( \hat{A}_1, \hat{A}_2,
\hat{A}_3, \cdots, \hat{A}_n, \cdots ) \rbrack =( \hat{A}_1,
\hat{A}_2, \hat{A}_3, \cdots, \hat{A}_n, \cdots ) \left(
\begin{array}{cccccc}
\lambda_1 & 0 & 0& & \cdots &\cdots\\ 0&  \lambda_2 & 0&  & \cdots
&\cdots\\ 0 & 0& \lambda_3 & & \cdots &\cdots\\ \vdots & \vdots &
\vdots & \ddots & \vdots & \vdots\\ 0 & 0 & 0& \cdots & \lambda_n
&\cdots\\ \vdots & \vdots & \vdots &  & \vdots & \vdots
\end{array}
\right), \eeq or
\begin{eqnarray*}
\lbrack  \hat{H},   \hat{A}_1 \rbrack &=& \hat{A}_1 \lambda_1\\
\lbrack  \hat{H},   \hat{A}_2 \rbrack &= &\hat{A}_2 \lambda_2\\
\cdots& \cdots & \cdots\\ \lbrack  \hat{H},   \hat{A}_n \rbrack&
=& \hat{A}_n \lambda_n\\ \cdots &\cdots&  \cdots.
\end{eqnarray*}

Thus, in general we get an unclosed algebra with infinitely many
shift operators. In case of the usual harmonic oscillator, we have
a pair of shift operators which can generate the entire spectrum.
In the present case, we have infinitely many pairs of shift
operators in which each pair is responsible for raising and
lowering the corresponding energy level. The whole spectrum can
only be generated by the infinite set of the shift operators
acting on the ground state. Hence, it is natural to see that all
the eigenvalues of the coefficient matrix are only dependent on
the energy of ground state. That is, $\lambda_i, (i=1, 2 , 3,
\cdots, n, \cdots)$ are the functions of the ground state energy.
Therefore, we can identify the operator $\hat{H}$ in the
coefficient matrix $T$ as the ground state energy! Based on this
observation, we can then get the analytic expression for the
energy of the ground state provided certain consistency relations
hold.

It remains to solve for the matrices $R, S $ and $ T$ given the
matrices $M_1, M_2, N_1$ and $N_2$. Once we have obtained the
matrices $R, S, T$ and diagonalize $T$, we effectively obtained
all the shift operators and we can then reconstruct the full
spectrum. It is instructive to note that one of the matrices $R$
or $S$ is redundant and we set it to unity. For convenience, let
us set $S$ in Eqs.(\ref{l10},\ref{l11}) to unity so that the
problem is reduce to the solution of $R, T$ from the following
equations
\begin{eqnarray}
\label{l12} M_1 R+M_2 &=&R T \\ \label{l13} N_1 R+N_2  &=&  T
\end{eqnarray}
and the diagonalization of $T$. Note that although we have
considered a specific form for the potential, the procedure
developed here is very general and can be applied to more
complicated cases.

\section{Analytic Expression for the Ground State
Energy: a Class of Anharmonic Potentials}\label{sect4}

To solve Eq.(\ref{l12}) and Eq.(\ref{l13}), we first make the
following observation. If we define

 \beq G=\left(
\begin{array}{cccccc}
1 & 0 & 0 & \cdots&\cdots&\cdots\\ 0 & 2 & 0 & \cdots
&\cdots&\cdots  \\ 0 & 0 & 3 & \cdots &\cdots&\cdots  \\ \vdots &
\vdots & \vdots & \ddots & \vdots& \vdots  \\ 0 &0 &0 & \cdots & n
& \cdots \\ \vdots & \vdots & \vdots & \vdots & \vdots & \ddots
\end{array}
\right), \eeq \beq P=\left(
\begin{array}{ccccccc}
0 &1 & 0 & 0 & \cdots&\cdots&\cdots\\ 0 &0 & 1 & 0 & \cdots
&\cdots&\cdots  \\ 0 &0 & 0 & 1 & \cdots &\cdots&\cdots  \\ \vdots
&\vdots & \vdots & \vdots & \ddots & \vdots& \vdots  \\ 0 &0 &0 &0
& \cdots & 1 & \cdots \\ \vdots &\vdots & \vdots & \vdots & \vdots
& \vdots & \ddots
\end{array}
\right) \;\;\;\;\;\; Q=\left(
\begin{array}{cccccc}
0 & 0 & 0 & \cdots&\cdots&\cdots\\ 1 & 0 & 0 &
\cdots&\cdots&\cdots\\ 0 & 1 & 0 & \cdots &\cdots&\cdots  \\ 0 & 0
& 1 & \cdots &\cdots&\cdots  \\ \vdots & \vdots & \vdots & \ddots
& \vdots& \vdots  \\ 0 &0 &0 & \cdots & 1 & \cdots \\ \vdots &
\vdots & \vdots & \vdots & \vdots & \ddots
\end{array}
\right),\eeq we have
\begin{equation}
\begin{array}{lll}
M_1& =& -PGPG\\ M_2&=&-2 \alpha G+2 \hat{H} PGP-2 \gamma QGQ
-\beta(2 G-1)Q\\ N_1&=&-2G\\ N_2&=&-GPGP.
\end{array} \label{eq33a}
\end{equation}
Here, the matrices $G, P$ and $Q$ are nothing but representations
of particle number, creation and annihilation operators for the
usual harmonic oscillator in Fock space. It is easy to check that
\begin{equation}
\label{l14} \lbrack  G, P   \rbrack = -P, \;\;\;\;\; \lbrack  G, Q
\rbrack = Q, \;\;\;\;\;\; PQ = 1,\;\;\;\;\;\; QP = diag(0, 1, 1,
\cdots, 1, \cdots).
\end{equation}
Furthermore, if we set $$ W=N_1 R, $$ then Eq.(\ref{l13}) becomes
$$ T=W+N_2 \hspace{0.8cm} R=N_1^{-1} W. $$  From Eq.(\ref{eq33a}),
we notice that $N_1 M_1 N_1^{-1}=N_2$, and as a consequence, we
can combine Eq.(\ref{l12}) and the above expression for the matrix
$T$, to yield the expression
\begin{equation} \label{l16} W^2+\lbrack W, N_2 \rbrack-N_1
M_2=0.
\end{equation}
Unfortunately, due to the infinite dimensionality of the matrices,
the solution, except in some special cases, is not known.

In the case of the one-dimensional oscillator with cubic-quartic
term, if we let $W$ assume the following form $$ W=2 (a G+b G Q+c
G P), $$ it is not difficult to show using Eq.(\ref{l14}) that

\begin{eqnarray}
W^2+\lbrack W, N_2 \rbrack - N_1 M_2 &= & 4 \{ (a^2+2 b c -\alpha)
G^2+ (a b -\frac{1}{2} \beta) (2 G^2-G) Q+(b^2 - \gamma)
GQGQ\nonumber \\ &&(a c + b) (2 G^2+G) P +(c^2 +a + \hat{H}) GPGP
\}
\end{eqnarray}
 so that consistency naturally
requires the following conditions
\begin{eqnarray}
 a^2+2 b c -\alpha&=& 0 \label{l19}\\ a b -\frac{1}{2} \beta&=&0\\
b^2 - \gamma&=&0\\ a c + b&=&0\\  c^2 +a + \hat{H}&=&0 \label{l17}
\end{eqnarray} in order that
Eq.(\ref{l16}) be satisfied. In particular, Eq.(\ref{l17}) gives
the ground state energy of the system.

As explained in the previous section, we have equated $\hat{H}$ to
one of its eigenvalues, namely the ground state energy. Thus, in
order that the full spectrum be generated from the infinitely many
raising operators acting on the ground state, so that each raising
operator generates its own corresponding energy level via its
action on the ground state, the consistency relations in
Eq.(\ref{l19})-Eq.(\ref{l17}) must hold.

It is easy to see that the solution to
Eq.(\ref{l19})-Eq.(\ref{l17}) is
\begin{equation}
\label{l31} a=\frac{\beta}{2 \sqrt{\gamma}} \;\;\;\;\;\;
b=\sqrt{\gamma} \;\;\;\;\;\; c=-2\frac{\gamma}{\beta}
\end{equation}
\begin{equation}
\hat{H}=-4\frac{\gamma^2}{\beta^2}-\frac{\beta}{2 \sqrt{\gamma}}
\end{equation}
\begin{equation}
\label{l20} \frac{\beta^2}{4 \gamma}-4
\frac{\gamma^{3/2}}{\beta}-\alpha=0
\end{equation}
and
\begin{equation}
\label{l32} a=-\frac{\beta}{2 \sqrt{\gamma}} \;\;\;\;\;\;
b=-\sqrt{\gamma} \;\;\;\;\;\; c=-2\frac{\gamma}{\beta}
\end{equation}
\begin{equation}
\hat{H}=-4\frac{\gamma^2}{\beta^2}+\frac{\beta}{2 \sqrt{\gamma}}
\end{equation}
\begin{equation}
\label{l21} \frac{\beta^2}{4 \gamma}+4
\frac{\gamma^{3/2}}{\beta}-\alpha=0.
\end{equation}

The additional constraint Eq.(\ref{l20}) (or Eq.(\ref{l21})) seems
to indicate that the ansatz for $W$ may be too simplistic.
However, we do not have a  direct solution of $W$ from
Eq.(\ref{l16}) for the given $N_1, N_2$ and $M_2$. Despite all
these, we can still obtain an analytic result for exact energy of
the ground state for the given anharmonic potential and this
result can prove to be valuable for analyzing the accuracy of
existing numerical methods. Besides, as mentioned before, it is
anticipated that the operator method can provide us with an
analytic tool for investigating the ground state\cite{raab},
something which is not possible using existing numerical
approaches.

To compute the energy levels of the excited states, we need to
diagonalize the following infinite dimensional matrix $$ T=\left(
\begin{array}{cccccccc}
2a  & 2c & -2 & 0 & 0 & & \cdots & \cdots\\ 4b & 4 a & 4c  & -6 &
0 && \cdots & \cdots \\ 0  & 6b  &6a & 6c & -12 && \cdots & \cdots
\\ 0 & 0 & 8b & 8a & 8c & \ddots & \cdots & \cdots \\ 0& 0& 0 &
10b & 10a & \ddots & -n(n+1) & \cdots \\ 0& 0& 0& 0 & 12b & \ddots
& 2nc & \cdots \\ 0& 0& 0& 0& 0 &  \ddots & 2na & \cdots \\ 0  & 0
& 0 & 0 & 0 &  & 2nb & \cdots \\ \vdots & \vdots & \vdots & \vdots
& \vdots &  & \vdots & \vdots
\end{array}
\right) $$ with $a, b$ and $c$ are given by Eq.(\ref{l31}) or
Eq.(\ref{l32}). Once we have diagonalized this matrix, using the
result of the ground state energy, all the other energy levels can
be obtained. Unfortunately, we do not know any method for
diagonalizing it at present moment despite the apparent simplicity
and symmetry in the matrix.

\section{Comparison of Numerical with the Analytic Results for
Ground State Energy}\label{sect5}

In this section, we use our analytic results to check against
previous numerical computations. For numerical simulation, we use
the method proposed by Ho {\it et al.} using the state-dependent
diagonalization method. As claimed in ref. \cite{lo}, this method
is very accurate and efficient compare to other numerical methods
for calculating the energy eigenvalues and eigenfunctions of the
one-dimensional harmonic oscillator with anharmonic potentials.

Due to the consistency requirements in Eq.(\ref{l20}) or
Eq.(\ref{l21}), the three coefficients $\alpha, \beta$ and
$\gamma$ in $$ V(x)=\alpha x^2+\beta x^3+\gamma x^4 $$ are not
mutually independent. It is not difficult to solve for $\beta$
using Eq.(\ref{l20}) or Eq.(\ref{l21}) for given values of
$\alpha$ and $\gamma$. The final solution is
\begin{equation}
\beta= 2 \sqrt{\gamma} \left[ \frac{\frac{\alpha}{3}}{
\left(\gamma+\sqrt{ \gamma^2-(\frac{\alpha}{3})^3 }
\right)^{\frac{1}{3}}} +\left(\gamma+\sqrt{
\gamma^2-(\frac{\alpha}{3})^3 } \right)^{\frac{1}{3}}  \right]
\label{f1}
\end{equation}
corresponding to Eq.(\ref{l20}) and $-\beta$ to Eq.(\ref{l21})
with the ground state energy being same in both cases.
Specifically, the ground state energy is given by \beq
E_0=-4\frac{\gamma^2}{\beta^2}-\frac{\beta}{2 \sqrt{\gamma}}.
\label{f2}\eeq The dependence of $\beta$ on $\alpha$ and $\gamma$
is shown in Fig.(\ref{fig1})

\bigskip

Setting $\alpha=1, -2$, the values of $E_0$ as a function of
$\gamma$ for the analytic formula and the numerical simulation
using the state dependent diagonalization \cite{lo} are shown
respectively  in Fig.(\ref{fig2}) and Fig.(\ref{fig3})) with the
corresponding potential for particular values of $\gamma$ as a
function of $x$ shown in Fig.(\ref{figpot})). The difference in
the energy computed from the numerical simulation and the analytic
formula as a function of $\gamma$ is also plotted in
Fig.(\ref{fig4}) and Fig.(\ref{fig5}) for $\alpha=1$ and
$\alpha=-2$ respectively. As shown in the figures, the numerical
simulation using state-dependent diagonalization is in excellent
agreement with our analytical formula. Moreover, we see that the
state-dependent-diagonalization method provides sufficiently high
accuracy for all practical purposes.

\section{Conclusions}\label{sect6}

In summary,we have formulated a general procedure for extracting
the full spectrum of a physical system with arbitrary potentials
using the operator method. We have applied it to solve the
one-dimensional harmonic oscillator with anharmonic problem. The
analytic expression for the ground state energy is obtained for a
large class of anharmonic potentials. Our results can be used to
verify the accuracy of existing numerical methods. However, in
order to get the full spectrum, we need to diagonalize an infinite
dimensional matrix. This last problem is to our knowledge an open
mathematical question except for some very special cases.

Our method has also confirmed the notion that once the Hamiltonian
of certain system is given, all the energy levels can be
determined through the operator method even without any prior
knowledge of a reference state, provided certain consistency
relations hold, a feat considered impossible hitherto.

Finally,  from Eq.(\ref{f1}) and Eq.(\ref{f2}), it easy to see
that in the limit when $\gamma \rightarrow \infty$, $E_0
\rightarrow -\frac{3}{2}(2 \gamma)^{1/3}$, confirming earlier
analysis on the mathematical properties of ground-state energy
obtained in ref. \cite{simo,ia}.

\section{Acknowledgment}

This work has been supported by NUS Research Grant No. RP3982713.
In addition, we would also like to thank the anonymous referee for
his invaluable remarks and comments.

\newpage

\begin{figure}[htb]
\caption{ The dependence of $\beta$ on  $\gamma$ and $\alpha$. }
\label{fig1}
\end{figure}

\begin{figure}[htb]
\caption{ The comparison of our analytic result (bold line) with
that obtained by state-dependent diagonalization method
(represented by the data points) for $\alpha=1$. } \label{fig2}
\end{figure}

\begin{figure}[htb]
\caption{ The comparison of our analytic result (bold line) with
that obtained by the state-dependent diagonalization method
(represented by data points) for $\alpha=-2$. } \label{fig3}
\end{figure}

\begin{figure}[htb]
\caption{ The potential, $V(x)$, as a function of $x$ for (a)
$\alpha=1$, $\gamma=1/2$ and (b) $\alpha=-2$, $\gamma=1/4$. }
\label{figpot}
\end{figure}

\begin{figure}[htb]
\caption{ The difference in the energy  between the analytic and
the numerical data as a function of $\gamma$ for $\alpha=1$. }
\label{fig4}
\end{figure}

\begin{figure}[htb]
\caption{ The difference in the energy  between the analytic and
the numerical data as a function of $\gamma$ for $\alpha=-2$.
}\label{fig5}
\end{figure}

\end{document}